\documentclass[aip,jcp,preprint]{revtex4-1}

\usepackage[pdftex]{graphics}
\usepackage{amssymb,amsmath}

\begin{document}

\title{An improved semiclassical theory of radical pair recombination reactions}
\author{D. E. Manolopoulos}
\affiliation{Department of Chemistry, University of Oxford, Physical and Theoretical Chemistry Laboratory, South Parks Road, Oxford, OX1 3QZ, UK}
\author{P. J. Hore}
\affiliation{Department of Chemistry, University of Oxford, Physical and Theoretical Chemistry Laboratory, South Parks Road, Oxford, OX1 3QZ, UK}
\begin{abstract}
We present a practical semiclassical method for computing the electron spin dynamics of a radical in which the electron spin is hyperfine coupled to a large number of nuclear spins. This can be used to calculate the singlet and triplet survival probabilities and quantum yields of radical recombination reactions in the presence of magnetic fields. Our method differs from the early semiclassical theory of Schulten and Wolynes [J. Chem. Phys. 68, 3292 (1978)] in allowing each individual nuclear spin to precess around the electron spin, rather than assuming that the hyperfine coupling-weighted sum of nuclear spin vectors is fixed in space. The downside of removing this assumption is that one can no longer obtain a simple closed-form expression for the electron spin correlation tensor: our method requires a numerical calculation. However, the computational effort increases only linearly with the number of nuclear spins, rather than exponentially as in an exact quantum mechanical calculation. The method is therefore applicable to arbitrarily large radicals. Moreover, it approaches quantitative agreement with quantum mechanics as the number of nuclear spins increases and the environment of the electron spin becomes more complex, owing to the rapid quantum decoherence in complex systems. Unlike the Schulten-Wolynes theory, the present semiclassical theory predicts the correct long-time behaviour of the electron spin correlation tensor, and it therefore correctly captures the low magnetic field effect in the singlet yield of a radical recombination reaction with a slow recombination rate. 
\end{abstract}

\maketitle

\section{Introduction}

Simulations of the quantum spin dynamics of coupled electron and nuclear spins are an essential component of many forms of electron paramagnetic resonance spectroscopy and are widespread in the field of spin chemistry. Used to interpret experimental data, to guide the design and execution of experiments, and to explore ideas not immediately amenable to experiment, such calculations are usually restricted to a total of $N\simeq 10$ spins, because the size of the spin Hilbert space increases exponentially with $N$.  But many -- probably the majority -- of paramagnetic species whose spin dynamics can be probed experimentally contain more than 10 nuclei with hyperfine coupling to one or more electron spins. For example, the radical pairs and triplet states responsible for magnetic field effects on the rates and yields of chemical reactions\cite{Steiner89,Rodgers09a,Hore12} and for spin hyperpolarization\cite{Woodward02,Matysik09} often have several tens of magnetic nuclei; the excitons, polaron pairs and bipolarons that appear to give rise to related magnetoresistance effects in polymeric organic semiconductors\cite{Koopmans11,Wohlgenannt12,Ehrenfreund12} may have $N$ approaching 100, and semiconductor quantum dots typically have $N > 10000$.\cite{Urbaszek13}

In spin chemistry, the simulation problem becomes particularly acute in the context of magnetoreception -- the proposal that photochemically produced radical pairs are responsible for the magnetic compass sense of migratory birds and possibly other animals.\cite{Schulten78b,Ritz00,Rodgers09b,Ritz11,Mouritsen12} The biological radicals thought to be involved here\cite{Maeda12} are multinuclear, long lived, slowly relaxing, and subject to an external magnetic field (the Earth's) that is neither much stronger nor much weaker than the majority of the anisotropic hyperfine interactions -- all factors that exacerbate the difficulty of spin dynamics simulations.

The upper limit on the $N$ that it is practical to simulate can sometimes be relaxed by exploiting special properties of particular spin systems, such as coupling topology, symmetry, conservation laws, and the existence of non-interacting sets of spin states.\cite{Hogben10} Experimental features can help too, for example by ensuring that a large fraction of the spin space is never populated and can therefore be removed from consideration.\cite{Hogben11} High-field, short-time, fast-relaxation and other approximations can also sometimes be useful. However, none of these tricks is generally applicable in the spin chemical context.

A potential solution to the exponential scaling problem was proposed by Schulten and Wolynes in the 1970s,\cite{Schulten78} when spin chemistry was still in its infancy. They described the electron spin motion induced by hyperfine coupling to many nuclear spins by means of a semiclassical  approximation in which the electron spin precesses around a hyperfine-weighted sum of nuclear spin vectors, which is assumed to be fixed in space. Applied to a radical pair derived from pyrene and {\em N,N}-dimethylaniline, the approach was shown to give excellent agreement with exact quantum simulations of the electron spin dynamics both with and without applied magnetic fields.\cite{Schulten78}

In this paper, we shall examine a straightforward extension to the Schulten-Wolynes theory in which each individual nuclear spin is allowed to precess around the electron spin. Similar extensions have been examined previously in the condensed matter physics (quantum dot) literature,\cite{Erlingsson04,Chen07} although in more sophisticated ways and with more sophisticated objectives than we shall pursue here (such as obtaining an analytical understanding of the asymptotic long-time decay of the electron spin polarisation). Our goal is simply to find a simulation method that scales linearly with $N$, approaches quantitative agreement with quantum mechanics as $N$ increases, and, unlike the Schulten-Wolynes theory, accurately captures the \lq low field effect'\cite{Brocklehurst76,Timmel98,Timmel04} of a weak external magnetic field on a slow radical pair recombination reaction. The motivation for this goal is that such a method would be very useful for simulating many of the current problems in spin chemistry.

\section{Theory}

\subsection{Radical pair recombination reactions}

To within a good approximation, the Hamiltonian that governs the rate of a radical pair recombination reaction in solution in the presence of an applied magnetic field is\cite{Schulten78}
$$
\hat{H} = \hat{H}_1+\hat{H}_2, \eqno(1)
$$
where
$$
\hat{H}_i = {\omega}_i\hat{S}_{iz}+\sum_{k=1}^{N_i} a_{ik}\hat{\bf I}_{ik}\cdot\hat{\bf S}_i. \eqno(2)
$$
Here ${\omega}_i=-\gamma_iB$, where $\gamma_i$ is the gyromagnetic ratio of the electron in radical $i$ and $B$ is the applied magnetic field, $N_i$ is the number of nuclear spins in the radical, $a_{ik}$ is an isotropic hyperfine coupling constant between the $k$-th nuclear spin and the electron spin, and $\hat{\bf I}_{ik}$ and $\hat{\bf S}_i$ are the corresponding nuclear and electron spin angular momentum operators. Note that this Hamiltonian neglects the Zeeman interactions between the nuclear spins and the magnetic field, which are typically far weaker than the terms that have been retained in Eq.~(2). We shall work throughout this paper in a unit system in which $\hbar=1$, in which the unit of time is the reciprocal of the unit of energy (or frequency, or magnetic field strength) that is used to specify $\omega_i$ and $a_{ik}$.

Suppose that the radical pair is produced photochemically in its singlet state. The initial density operator will then be
$$
\hat{\rho}(0) = {1\over Z_1Z_2}\hat{P}_{\rm S}, \eqno(3)
$$
where $Z_i=\prod_{k=1}^{N_i} (2I_{ik}+1)$ is the total number of nuclear spin states in radical $i$ and
$$
\hat{P}_{\rm S}= {1\over 4}-\hat{\bf S}_1\cdot\hat{\bf S}_2 \eqno(4)
$$
is the projection operator onto the singlet electronic subspace. As time evolves, the singlet and triplet states will be mixed by the Zeeman and hyperfine interactions in each radical, so the probability of finding the radical pair in the singlet state will decrease from one. This singlet probability is given at time $t$ by
$$
P_{\rm S}(t) = {\rm tr}\left[\hat{\rho}(t)\hat{P}_{\rm S}\right], \eqno(5)
$$
where
$$
\hat{\rho}(t) = e^{-i\hat{H}t}\hat{\rho}(0)\,e^{+i\hat{H}t}. \eqno(6)
$$
Combining these equations with Eq.~(1), and noting that $\hat{H}_1$ and $\hat{H}_2$ commute, one finds that $P_{\rm S}(t)$ can be written equivalently as\cite{Schulten78}
$$
P_{\rm S}(t) = {1\over 4}+\sum_{\alpha\beta} R^{(1)}_{\alpha\beta}(t)R^{(2)}_{\alpha\beta}(t), \eqno(7)
$$
where
$$
R^{(i)}_{\alpha\beta}(t) = {1\over Z_i}{\rm tr}\left[\hat{S}_{i\alpha}\,e^{+i\hat{H}_it}\hat{S}_{i\beta}\,e^{-i\hat{H}_it}\right] = {1\over Z_i}{\rm tr}\left[\hat{S}_{i\alpha}(0)\hat{S}_{i\beta}(t)\right] \eqno(8)
$$
is an electron spin correlation tensor for the spin in radical $i$. The elements of this tensor can be shown to be real, and to satisfy
$$
R^{(i)}_{xx}(t) = R^{(i)}_{yy}(t), \eqno(9)
$$
$$
R^{(i)}_{yx}(t) = -R^{(i)}_{xy}(t), \eqno(10)
$$
$$
R^{(i)}_{xz}(t) = R^{(i)}_{yz}(t) = R^{(i)}_{zx}(t) = R^{(i)}_{zy}(t) = 0. \eqno(11)
$$
The problem of calculating the time-dependent singlet probability $P_{\rm S}(t)$ is thus reduced to one of calculating the $xx$, $xy$, and $zz$ components of the electron spin correlation tensor for each radical separately.

In an experiment one typically does not measure the singlet probability $P_{\rm S}(t)$ directly, but rather the overall quantum yield $\Phi_{\rm S}$ of the singlet product. This is usually assumed to be given by a simple exponential model\cite{Brocklehurst96}
$$
\Phi_{\rm S}(B) = k\int_0^{\infty} P_{\rm S}(t)e^{-kt}\,dt, \eqno(12)
$$
in which $k$ is a first order rate constant for radical pair recombination. $\Phi_{\rm S}$ depends on the magnetic field by virtue of the $\omega_i{S}_{iz}$ term in Eq.~(2), and there are two separate magnetic field effects that are of interest. In a high magnetic field the electronic Zeeman splitting is so large that the ${\rm T}_{\pm 1}$ components of the triplet state are energetically inaccessible from the singlet, so ${\rm S}$ can only mix with ${\rm T}_0$ and the singlet yield is enhanced relative to its field-free value.\cite{Steiner89} Conversely, a low but non-zero magnetic field breaks the symmetry of the zero-field problem and splits the degeneracy of the zero-field eigenstates, producing more pathways for (energetically feasible) singlet to triplet interconversion and lowering the singlet quantum yield from its field-free value.\cite{Brocklehurst76,Brocklehurst96,Timmel98,Till98}

\subsection{Schulten-Wolynes theory}

The Schulten-Wolynes theory is based on the assumption that, in a radical with a sufficiently large number of nuclear spins, the details of the nuclear spin dynamics will cease to matter: the electron spin in the radical will simply see a hyperfine-weighted sum of nuclear spin vectors, which can be regarded as being fixed in space. The electron spin will then precesses around the resultant of this vector and the applied magnetic field.\cite{Schulten78}

Consistent with the assumption of a large number of nuclear spins, Schulten and Wolynes calculate the distribution of the hyperfine-weighted sum of nuclear spin vectors using the asymptotic ($N_i\to\infty$) result for the end-to-end distribution of a random flight polymer.\cite{Flory69} If 
$$
{\bf I}_i = \sum_{k=1}^{N_i} a_{ik}{\bf I}_{ik}, \eqno(13)
$$
and each ${\bf I}_{ik}$ is interpreted as a classical vector of length $\sqrt{I_{ik}(I_{ik}+1)}$, this is\cite{Schulten78}
$$
P({\bf I}_i) = \left({\tau_i^2\over 4\pi}\right)^{3/2}e^{-I_i^2\tau_i^2/4}, \eqno(14)
$$
where $I_i=|{\bf I}_i|$ and
$$
\tau_i^2 = {6\over \sum_k a_{ik}^2I_{ik}(I_{ik}+1)}. \eqno(15)
$$
For each ${\bf I}_i$ in this distribution, the electron spin in radical $i$ is assumed to precess around the vector
$$
\boldsymbol{\omega} = \boldsymbol{\omega}_i+{\bf I}_i, \eqno(16)
$$
where $\boldsymbol{\omega}_i^{\rm T}=(0,0,\omega_i)$ accounts for the applied magnetic field. The contribution that this precession makes to the electron spin correlation tensor of the radical can be worked out either quantum mechanically or semiclassically. Schulten and Wolynes chose to do the calculation quantum mechanically,\cite{Schulten78} but a semiclassical calculation gives the same result.

Consider the contribution
$$
R^{(i)}_{\alpha\beta}(t;\boldsymbol{\omega}) = {\rm tr}_S\left[\hat{S}_{i\alpha}(0)\hat{S}_{i\beta}(t)\right]\eqno(17)
$$
to the electron spin correlation tensor that comes from the precession of the electron spin around a particular vector $\boldsymbol{\omega}$ in Eq.~(16), where we have used ${\rm tr}_S$ to denote a trace over the electron spin states. The simplest possible semiclassical (classical vector model) approximation to this is
$$
R^{(i)}_{\alpha\beta}(t;\boldsymbol{\omega}) \simeq {2S+1\over 4\pi}\int d\Omega_S\,S_{i\alpha}(0)S_{i\beta}(t), \eqno(18)
$$
in which the quantum mechanical trace has been replaced by an integral 
$$
{\rm tr}_S\to {2S+1\over 4\pi}\int d\Omega_S \equiv {2S+1\over 4\pi}\int_0^{2\pi} d\phi\int_0^{\pi}\sin\theta\,d\theta \eqno(19)
$$
over the orientation of a classical electron spin vector
$$
{\bf S}_i^{\rm T}(0) = \sqrt{S(S+1)}(\sin\theta\cos\phi,\sin\theta\sin\phi,\cos\theta), \eqno(20)
$$
with $S=1/2$. The precession of this vector
$$
{d\over dt}{\bf S}_i(t) = \boldsymbol{\omega} \times{\bf S}_i(t), \eqno(21)
$$
gives
$$
{\bf S}_i(t) = {\bf S}_i^{\parallel}(0)+{\bf S}_i^{\perp}(0)\cos\omega t + {\bf S}_i^{\times}(0)\sin\omega t,
\eqno(22)
$$
where $\omega=|\boldsymbol{\omega}|$ and
$$
{\bf S}_i^{\parallel}(0)=\hat{\boldsymbol{\omega}}\hat{\boldsymbol{\omega}}^{\rm T}{\bf S}_i(0), 
$$
$$
{\bf S}_i^{\perp}(0) = \left[{\boldsymbol{1}}-\hat{\boldsymbol{\omega}}\hat{\boldsymbol{\omega}}^{\rm T}\right]{\bf S}_i(0),
$$
$$
{\bf S}_i^{\times}(0)=\hat{\boldsymbol{\omega}}\times {\bf S}_i(0), 
\eqno(23)
$$
with $\hat{\boldsymbol{\omega}}=\boldsymbol{\omega}/\omega$. Substituting Eq.~(22) into Eq.~(18), and doing the integrals over $\theta$ and $\phi$ in Eq.~(19), one finds that
$$
R^{(i)}_{\alpha\beta}(t;\boldsymbol{\omega}) = {S(S+1)(2S+1)\over 3}\left[\hat{\omega}_{\alpha}\hat{\omega}_{\beta}+(\delta_{\alpha\beta}-\hat{\omega}_{\alpha}\hat{\omega}_{\beta})\cos\omega t + \sum_{\gamma} \epsilon_{\alpha\beta\gamma}\hat{\omega}_{\gamma}\sin\omega t\right], \eqno(24)
$$
where $\delta_{\alpha\beta}$ is the Kronecker delta and $\epsilon_{\alpha\beta\gamma}$ is the alternating tensor. This is the correct quantum mechanical result: the semiclassical approximation in Eq.~(18) is {\em exact} for the correlation tensor of an isolated spin in the presence of a fixed magnetic field.

The final stage of the calculation is to average $R^{(i)}_{\alpha\beta}(t;\boldsymbol{\omega})$ in Eq.~(24) over the distribution of ${\bf I}_i$ in Eq.~(14). It is most convenient to do this by re-writing the distribution as 
$$
P(\boldsymbol{\omega}) = \left({\tau_i^2\over 4\pi}\right)^{3/2}e^{-(\omega^2+\omega_i^2-2\omega\omega_i\cos\theta)\tau_i^2/4}, \eqno(25)
$$
so that the Schulten-Wolynes approximation to $R_{\alpha\beta}^{(i)}(t)$ becomes
$$
R^{(i)}_{\alpha\beta}(t) \simeq \int_0^{2\pi} d\phi \int_{0}^{\pi} \sin\theta d\theta \int_0^{\infty} \omega^2d\omega\, P(\boldsymbol{\omega})R^{(i)}_{\alpha\beta}(t;\boldsymbol{\omega}), \eqno(26)
$$
where $\theta$ and $\phi$ are now the polar and azimuthal angles of the vector $\boldsymbol{\omega}$. Rodgers\cite{Rodgers07} has shown that when Eqs.~(24) and~(25) are substituted into Eq.~(26), all three integrals can be evaluated in closed form. Setting $S=1/2$ for the electron spin, the final results for the three independent components of the tensor are\cite{Rodgers07}
$$
R^{(i)}_{xx}(t) = \left[\omega_*\left(2+e^{-t_*^2}\left[(\omega_*^2-2)\cos \omega_*t_*-2\omega_*t_*\sin\omega_*t_*\right]\right)-4f(\omega_*,t_*)\right]/2\omega_*^3,
$$
$$
R^{(i)}_{xy}(t) = e^{-t_*^2}\left[2\omega_*t_*\cos\omega_*t_*+(\omega_*^2-2)\sin\omega_*t_*\right]/2\omega_*^2,
$$
$$
R^{(i)}_{zz}(t) = \left[\omega_*(\omega_*^2+4e^{-t_*^2}\cos \omega_*t_*-4)+8f(\omega_*,t_*)\right]/2\omega_*^3, \eqno(27)
$$
where $\omega_*=\omega_i\tau_i$, $t_*=t/\tau_i$, and the one remaining integral
$$
f(\omega_*,t_*) = \int_0^{t_*} e^{-s_*^2}\sin\omega_*s_*\, ds_* \eqno(28)
$$
can be expressed in terms of complex error functions.

\subsection{Improved semiclassical theory}

The assumptions of the Schulten-Wolynes theory are certainly likely to become valid at sufficiently short times as the number of nuclear spins in the radical tends to infinity. However, it is not clear that they will be so well justified for a radical containing several tens of nuclear spins, which is typical for the radicals of interest in organic radical recombination reactions. In particular, while the assumption of an asymptotic (Gaussian) distribution for the sum of hyperfine-weighted nuclear spin vectors might be reasonable for such a radical, the assumption that this vector remains fixed in space during the electron spin evolution may well not be. (The argument that supports neglecting the nuclear spin precession is that in a radical with $N_i$ nuclear spins the precession frequency of the electron spin will be on the order of $\sqrt{N_i}a$, whereas that of each nuclear spin will be on the order of $a$, where $a$ is the average hyperfine coupling constant.\cite{Erlingsson02} For sufficiently large $N_i$, the nuclear spin precession will therefore be much slower than the electron spin precession, and so from the point of view of the electron spin dynamics the nuclear spins can be regarded as constant. While this approximation may well be valid for moderate times for the electron spin in a quantum dot ($N_i>10^4$), it is not at all clear that the separation of time scales will be large enough to justify it when $N_i$ is only a small multiple of ten.)

Fortunately, both assumptions can be removed by modifying the Schulten-Wolynes theory in a fashion suggested by the semiclassical approximation in Eq.~(18). Suppose we make the same approximation to the traces over the nuclear spins in the radical as we have made for the electron spin,
$$
{\rm tr}_{I_{ik}} \to {(2I_{ik}+1)\over 4\pi} \int d\Omega_{I_{ik}}, \eqno(29)
$$
so that each nuclear spin is regarded as a classical vector of length $\sqrt{I_{ik}(I_{ik}+1)}$ with an initial orientation that is distributed uniformly on the surface of a sphere. Then it should be clear from Eq.~(18) that we can approximate the spin correlation tensor $R^{(i)}_{\alpha\beta}(t)$ by
$$
R^{(i)}_{\alpha\beta}(t) \simeq {1\over 2\pi} \int d\Omega_S\,
\prod_{k=1}^{N_i} {1\over 4\pi} \int d\Omega_{I_{ik}}\, S_{i\alpha}(0)S_{i\beta}(t), \eqno(30)
$$
where we have explicitly set $S=1/2$ and cancelled the factors of $(2I_{ik}+1)$ in Eq.~(29)
with the factor of $1/Z_i$ in Eq.~(8). The key difference between this and the Schulten-Wolynes theory is that each nuclear spin in the radical is now an independent vector which is free to precess around the electron spin at the same time as the electron spin precesses around it. That is, we can use the dynamically coupled equations of motion  
$$
{d\over dt}{\bf S}_i(t) = \left[\boldsymbol{\omega}_i+\sum_{k=1}^{N_i} a_{ik}{\bf I}_{ik}(t)\right]\times {\bf S}_i(t), \eqno(31)
$$
and
$$
{d\over dt}{\bf I}_{ik}(t) = a_{ik}{\bf S}_i(t) \times {\bf I}_{ik}(t),  \eqno(32)
$$
to obtain the components $S_{i\beta}(t)$ of ${\bf S}_i(t)$ in Eq.~(30), thereby eliminating the second of the assumptions made in the Schulten-Wolynes theory. 

These modifications clearly preclude obtaining a simple closed form expression for $R^{(i)}_{\alpha\beta}(t)$: Eqs.~(30) to (32) have to be solved numerically.\cite{Yuzbashyan05} However, this is not especially difficult. The integrals over $d\Omega_S$ and $d\Omega_{I_{ik}}$ in Eq.~(30) can be done by Monte Carlo integration by sampling the initial electron and nuclear spin vectors at random from the surfaces of their respective spheres. The coupled equations of motion in Eqs.~(31) and~(32) could in principle be solved using any one of a number of ordinary differential equation integrators, although we prefer to use a specialised integrator for this purpose as described in Appendix A. No matter what integrator one uses, the number of coupled equations that must be integrated increases only linearly with $N_i$, so the calculation is entirely feasible for a radical with any number of nuclear spins.

\section{Results and Discussion}

\subsection{A well-studied radical pair}

As a first test of this modified semiclassical theory, we have used it to calculate the time-dependent triplet probability $P_{\rm T}(t)=1-P_{\rm S}(t)$ of the pyrene + {\em N,N}-dimethylaniline ($^2$Py$^-$ + $^2$DMA$^+$) radical pair that was considered by Schulten and Wolynes\cite{Schulten78} and subsequently by Knapp and Schulten.\cite{Knapp79} The spin dynamics of this radical pair had previously been solved quantum mechanically by Werner {\em et al.}\cite{Werner77} We have adopted the same (simplified) set of hyperfine coupling constants as in these earlier studies, namely $4\times (a_{\rm H}=0.23\hbox{ mT})+4\times (a_{\rm H}=0.52\hbox{ mT})$ for $^2$Py$^-$ and $6\times (a_{\rm CH3}=1.2\hbox{ mT})+3\times (a_{\rm H}=0.625\hbox{ mT})+1\times(a_{\rm N}=1.2\hbox{ mT})$ for $^2$DMA$^+$. To facilitate a direct comparison with the results reported in Refs.~\cite{Schulten78,Knapp79,Werner77}, we have also converted the unit of time from mT$^{-1}$ to ns using the gyromagnetic ratio of a free electron ($\gamma_{\rm e}=0.176$ ns$^{-1}$mT$^{-1}$).

Fig.~1 shows the resulting triplet probabilities for a $^2$Py$^-$ + $^2$DMA$^+$ radical pair initially formed in the singlet state, for various magnetic field strengths. The exact quantum mechanical (QM) results are compared with those of the Schulten-Wolynes (SW) theory and our modified semiclassical (SC) approximation. One sees that both semiclassical theories do remarkably well in capturing the overall behaviour of the QM triplet probability, especially at the highest magnetic field strength considered in the figure (8 mT, almost an order of magnitude larger than the largest hyperfine coupling constant in the radical pair). However, the present SC theory gives a somewhat more accurate result than the SW theory even at this field strength, and the difference between the two approximations becomes more pronounced as the field strength is decreased. For this problem, the SW approximation clearly works very well. However the present SC approximation is noticeably more accurate  when the field strength is comparable to the hyperfine coupling constants in $^2$DMA$^+$ ($\sim 1$ mT), and also in the limit of zero applied magnetic field.

\subsection{Electron spin correlation tensors}

\begin{table}[Ht]
\begin{center}
\caption{Hyperfine coupling constants used in the calculations reported in Figs.~2 to~5.}
\bigskip
\begin{tabular}{lr} \hline \hline
$k$ & $a_{ik}$ (mT)\\
\hline
     1 & $-$0.999985 \\
     2 & $-$0.7369246 \\
     3 & 0.511210 \\
     4 & $-$0.0826998 \\
     5 & 0.0655341 \\
     6 & $-$0.562082 \\
     7 & $-$0.905911 \\
     8 & 0.357729 \\
     9 & 0.358593 \\
    10 & 0.869386 \\
    11 & $-$0.232996 \\
    12 & 0.0388327 \\
    13 & 0.661931 \\
    14 & $-$0.930856 \\
    15 & $-$0.893077 \\
    16 & 0.0594001 \\
\hline \hline
\end{tabular}
\end{center}
\end{table}

Considerably more insight into the performance of the SW and SC approximations can be obtained by using them to compute the \lq raw' electron spin correlation tensors of a series of radicals of increasing size containing $I=1/2$ nuclei with hyperfine coupling constants distributed uniformly in the range $-1\le a_{ik}\le 1$ mT. The hyperfine constants that we actually used in these calculations were obtained from a random number generator and are listed for completeness in Table~I.

Fig.~2 shows the resulting electron spin correlation tensors for a magnetic field strength of 0.5 mT, which lies in the middle of the range of $|a_{ik}|$ values and so probes the regime in which the Zeeman and hyperfine interactions are of a comparable magnitude and contribute equally to the electron spin dynamics. The QM, SW, and SC results are compared for radicals with 1, 4 and 16 nuclear spins. In the smallest of these radicals, the components of the QM electron spin correlation tensor exhibit coherent oscillations which are not captured by either semiclassical theory. However, the present SC method clearly does a better job of following the average of the quantum oscillations than the SW theory, especially at long times. When there are 4 nuclear spins in the radical, the amplitude of the coherent quantum oscillations is significantly suppressed, and by the time there are 16 nuclear spins the oscillations are washed out almost completely: all that is left is the average classical precessional behaviour, which is captured almost perfectly by the present SC theory. This is simply a consequence  of the quantum decoherence that one expects to see in strongly-coupled multi-dimensional systems.\cite{Miller12}

We have also performed calculations for a variety of other magnetic field strengths and obtained results consistent with those shown in Fig.~2. In the high-field limit, both the original SW theory and our modified SC theory become exact, for the reasons explained following Eq.~(24) -- the classical vector model gives the exact result for precession around a fixed field. That neither is exact in the low-field limit can be seen from Fig.~3, which shows how the electron spin correlation tensors behave in the absence of a magnetic field. One again sees that, while neither semiclassical method captures the coherent oscillations in the QM electron spin correlation tensor of a radical with just one nuclear spin, the present SC theory becomes almost quantitatively accurate by the time the radical contains 16 nuclear spins. This is not the case for the SW theory: the components of the electron spin correlation tensor in Eq.~(27) become constant beyond $t\simeq 2\tau_i$ and fail to capture the long-time behaviour of the electron spin dynamics.

The fact that the present SC theory begins to agree with quantum mechanics in a radical with 16 nuclear spins is rather fortunate, because this is close to the maximum number of spins that can be included in an exact QM calculation. In a radical with 16 spin-1/2 nuclei, the Hilbert space contains 2$^{17}=131,072$ state vectors, and the QM calculation is already approaching the limit of practical feasibility. This is illustrated in Fig.~4, which compares the relative computer times of the SC and QM calculations for radicals containing an increasing number of nuclear spins. For a radical with fewer than 10 nuclear spins, the SC calculation is significantly slower than the QM calculation, because of the need to run a large number of classical trajectories to converge the Monte Carlo integration over the initial spin orientations in Eq.~(30).\cite{MonteCarlo} However, since the computational effort in the SC case increases only linearly with the number of nuclear spins, whereas in the QM case it increases exponentially, the SC calculation must eventually become more efficient, and in our implementation the cross-over occurs at 12 nuclear spins. 

\subsection{The low field effect}

In view of the results in Figs.~2 and 3, it should be clear that one circumstance in which the present SC theory is likely to do significantly better than the SW theory is in capturing the low field effect in the singlet yield of a radical pair recombination reaction with a slow recombination rate. In order to investigate this, we have calculated the singlet yield in Eq.~(12) for a radical pair with no nuclear spins in one radical and 12 $I=1/2$ nuclear spins in the other, as a function of the magnetic field strength $B$ and the recombination rate constant $k$. The reason for taking one of the radicals in the pair to have no hyperfine interactions and the other to have many is that this has been established experimentally to be the situation in which the low field effect is most pronounced.\cite{Henbest06,Rodgers07b} The reason for stopping at 12 $I=1/2$ nuclear spins in the large radical is simply one of convenience: these calculations have to be done for a range of magnetic field strengths and for sufficiently long times to ensure the convergence of the integral in Eq.~(12), and so are computationally more demanding than any of the calculations we have reported so far. While this is not an issue for the semiclassical (SW and SC) methods, it is an issue for the QM method that is used to generate exact results for comparison.

Fig.~5 compares the low field effects obtained from the QM, SC and SW theories for this idealised radical pair, the hyperfine coupling constants in the larger radical being distributed uniformly in the range $-1\le a_{ik}\le 1$ mT. The ordinate of the plot is the difference between the singlet yield at magnetic field $B$ and the singlet yield in the absence of a field, $\Phi_{\rm S}(B)-\Phi_{\rm S}(0)$, which is accessible experimentally as the integrated signal obtained from a modulated MARY (Magnetically Affected Reaction Yield) experiment.\cite{Hamilton88,Batchelor93} The three panels of the figure show the results for three different recombination rates, between 0.05 and 0.15 mT, all of which are smaller than the average hyperfine coupling in the larger radical. With these parameters, the low field effect in the exact QM calculation is seen to be quite pronounced, but not unreasonable: low field effects of this magnitude have certainly been observed experimentally.\cite{Henbest06}

As one would expect on the basis of the electron spin correlation tensors in Figs.~2 and~3, the present SC theory does a remarkably good job of capturing the correct low field effect, even in a radical pair with as few as 12 nuclear spins. The SC and QM results are in almost quantitative agreement for all three radical pair recombination rates. However, while it is qualitatively correct in terms of its dependence on the magnetic field strength, the low field effect predicted by the SW theory for this radical pair is far too large -- by a factor of almost three at the lowest radical pair recombination rate we have considered ($k=0.05$ mT). Clearly this is because the SW theory does not capture the correct long-time behaviour of the electron spin dynamics, which contributes more to the singlet yield as the recombination rate constant decreases. This final numerical example therefore provides a clear illustration of the limitations of the SW approximation and the utility of our modification to it.

\subsection{Discussion}

One important thing we have not been able to establish with the above numerical examples is whether the present SC theory will become even more accurate as the number of nuclear spins in the radical increases further. However, we do believe that this will be the case on the basis of some recent results from the condensed matter physics literature.\cite{Erlingsson04,Chen07}

The electron spin dynamics generated by the Hamiltonian $H_i$ in Eq.~(2) is known in that literature as the \lq central spin problem'. This problem has been studied extensively during the last decade because of its relevance to the electron spin dynamics in GaAs quantum dots, which contain significantly more ($10^4-10^6$) nuclear spins than the radicals of interest in radical pair recombination reactions. Early theoretical work on the problem stressed the importance of quantum coherence in the electron spin dynamics and advocated a fully quantum mechanical treatment,\cite{Khaetskii02,Khaetskii03} which is clearly only possible in certain limiting cases (such as the limit of a large applied magnetic field). However, it has since been established quite generally, both on physical grounds and more rigorously, that the classical equations of motion will become exact for measurements of the electron spin dynamics in a radical with sufficiently many nuclear spins.\cite{Chen07}

We find it especially interesting that a fully quantum mechanical treatment was initially thought to be important for studying the electron spin dynamics in quantum dots.\cite{Khaetskii02,Khaetskii03}  The same can be said for the low field effect in radical pair recombination reactions. Although simple classical vector arguments have provided useful insights into this effect,\cite{Brocklehurst76,Till98} it is probably true to say that the majority of the existing literature attributes it to quantum coherences arising from the off-diagonal elements of the electron spin density matrix.\cite{Timmel98} But as we have shown in Figs. 2 and~3, these coherences are rapidly quenched in a radical with more than a handful of nuclear spins: all that survives is the average classical precessional behaviour, which accounts almost quantitatively for the low field effect in a radical with as few as 12 spins (see Fig.~5).

\section{Concluding remarks}

From the perspective of simulating spin chemistry with which we began in the Introduction, the results in Figs.~1-5 are clearly very encouraging. The present SC approximation misses the quantum coherence that is present in the electron spin dynamics of a radical with fewer than 10 or so nuclear spins, but we feel that this is largely irrelevant. If one wants to simulate such a radical, one can always do so quantum mechanically (see Fig.~4). And as soon as the number of nuclear spins increases to the point where the exact QM calculation becomes impractical, the accuracy of the present SC approximation becomes almost quantitative (see Figs.~2, 3, and 5). 

The present results therefore open up an exciting avenue for future research. Because the SC theory is so simple, being based entirely on the primitive classical vector model, it is straightforward to see how to generalise it to include many of the effects that are not included  in the zeroth-order spin Hamiltonian in Eq.~(1), such as anisotropic hyperfine interactions, exchange and dipolar interactions between the electrons in a radical pair, spin relaxation processes, and so on. To give just one example of the many possible applications of the theory, it would be straightforward to use it to investigate potential low field effect-enhancing mutations of the carotenoid-porphyrin-fullerene triad that has recently been established as a proof-of-principle for the operation of a chemical compass.\cite{Maeda08} Such an investigation would be totally impractical using exact quantum mechanical methodology, because the carotenoid radical in the photo-excited triad contains $\sim 46$ protons with hyperfine couplings larger than the Earth's magnetic field, implying a Hilbert space of $\sim 10^{14}$ states. With the present SC theory, this investigation would be entirely feasible, and the results we have presented here suggest that its conclusions would be perfectly reasonable.

\begin{acknowledgements} 
We are indebted to David Reichman for pointing us to work on the central spin problem in the condensed matter physics community, and to Christopher Rodgers for his unpublished work on the SW approximation (Ref.~\cite{Rodgers07}). DEM acknowledges support from the Wolfson Foundation and the Royal Society and PJH acknowledges funding from DARPA (QuBE: N66001-10-1-4061). 
\end{acknowledgements}

\section{Appendix}
\subsection{Classical evolution}

The algorithm we have used to integrate the coupled equations of motion in Eqs.~(31) and~(32) through a time step $\delta t$ can be summarised as follows:
\begin{enumerate}
\item
Let $\boldsymbol{\omega} = \boldsymbol{\omega}_i+\sum_{k=1}^{N_i} a_{ik}{\bf I}_{ik}$, and, regarding this as a constant vector, use Eqs.~(22) and~(23) to evolve ${\bf S}_i(t)$ for time $\delta t/2$.
\item
For $k=1,\ldots,N_i$, let $\boldsymbol{\omega} = a_{ik}{\bf S}_i$, and, regarding this as a constant vector, use Eqs.~(22) and~(23) with ${\bf S}_i\to{\bf I}_{ik}$ to evolve ${\bf I}_{ik}(t)$ for time $\delta t$. 
\item
Let $\boldsymbol{\omega} = \boldsymbol{\omega}_i+\sum_{k=1}^{N_i} a_{ik}{\bf I}_{ik}$, and, regarding this as a constant vector, use Eqs.~(22) and~(23) to evolve ${\bf S}_i(t)$ for time $\delta t/2$.
\end{enumerate}
This algorithm is only second order accurate, giving an error of $O(\delta t^3)$ per time step and an error of $O(\delta t^2)$ for propagation over a fixed time interval. However it is easy to program, has minimal storage requirements, and exactly conserves both the lengths of all (electron and nuclear) spin vectors and the classical energy $\boldsymbol{\omega}_i\cdot{\bf S}_i+\sum_{k=1}^{N_i} a_{ik}{\bf I}_{ik}\cdot {\bf S}_i$. One constant of the motion that it does not exactly conserve is the projection of the total (electronic and nuclear spin) angular momentum along $\boldsymbol{\omega}_i$, which can therefore be used to check the accuracy of the integration. 

\subsection{Quantum evolution}

The $xx$, $xy$ and $zz$ components of the quantum mechanical electron spin correlation tensor in Eq.~(8) can be computed as follows. Suppose that, for each composite nuclear spin state $\left|M\right>=\left|M_{I_{i1}},\ldots,M_{I_{iN_i}}\right>$, we propagate the pair of state vectors $\left|\sigma M\right>$ with $\sigma=\,\uparrow$  for $M_S=+{1\over 2}$ and $\downarrow$ for $M_S=-{1\over 2}$ forwards in time to obtain their time-dependent expansion coefficients 
$$
U_{\sigma' M',\sigma M}(t) = \left<\sigma' M'\right|e^{-i\hat{H}_it}\left|\sigma M\right>. \eqno(B1)
$$
Then it follows from Eq.~(8) and the properties of the electron spin operators that
$$
R^{(i)}_{xx}(t)+iR^{(i)}_{xy}(t) = {1\over 2Z_i}\sum_{M',M} \Bigl[U_{\uparrow M',\uparrow M}(t)^*U_{\downarrow M',\downarrow M}(t)+U_{\uparrow M',\downarrow M}(t)^*U_{\downarrow M',\uparrow M}(t)\Bigr], \eqno(B2)
$$ 
and
$$
R^{(i)}_{zz}(t) = {1\over 4Z_i}\sum_{M',M}\Bigl[\left|U_{\uparrow M',\uparrow M}(t)\right|^2-\left|U_{\uparrow M',\downarrow M}(t)\right|^2 - \left|U_{\downarrow M',\uparrow M}(t)\right|^2+ \left|U_{\downarrow M',\downarrow M}(t)\right|^2\Bigr]. \eqno(B3)
$$
The required components of $R_{\alpha\beta}^{(i)}(t)$ can therefore be obtained by propagating $Z_i$ independent pairs of state vectors forwards in time one pair at a time, and accumulating their expansion coefficients in the basis $\left|\sigma' M'\right>$ using these equations. 

The time evolution of the pair of state vectors is conveniently performed with a symplectic  integrator.\cite{Gray96} This propagates each state vector in the pair
$$
\left|\psi(t)\right> = e^{-i\hat{H}_it}\left|\sigma M\right> = \left|q(t)\right>+i\left|p(t)\right>, \eqno(B4)
$$  
through a time interval $\delta t$ using the following $m$-step algorithm (for $j=1,\ldots,m$),
$$
\left|p_{j}\right> = \left|p_{j-1}\right>-b_j\delta t \hat{H}_i\left|q_{j-1}\right>, \eqno(B5)
$$
$$
\left|q_{j}\right> = \left|q_{j-1}\right>+a_j\delta t \hat{H}_i\left|p_j\right>, \eqno(B6)
$$
where $\left|p_0\right>=\left|p(t)\right>$ and $\left|q_0\right>=\left|q(t)\right>$ give $\left|p_m\right>\simeq \left|p(t+\delta t)\right>$ and $\left|q_m\right>\simeq \left|q(t+\delta t)\right>$ with an error of $O(\delta t^{n+1})$. Note that these equations only involve Hamiltonian matrix-state vector multiplications, which can be facilitated by exploiting the sparsity of the matrix representation of $\hat{H}_i$: each of the $2Z_i$ basis states is coupled to at most $N_i\sim \log Z_i$ others by the Hamiltonian in Eq.~(2). Exploiting this sparsity also leads to an algorithm with a minimal storage requirement, which is particularly useful for radicals with a large number of nuclear spins. In practice, we have found that the four-stage, fourth-order ($m=n=4$) symplectic integrator coefficients $a_j$ and $b_j$ given in Ref.~\cite{Gray96} provide an ideal balance between accuracy and efficiency for the present problem. 

\begin{figure}[Ht]
\centering
\resizebox{0.8\columnwidth}{!} {\includegraphics{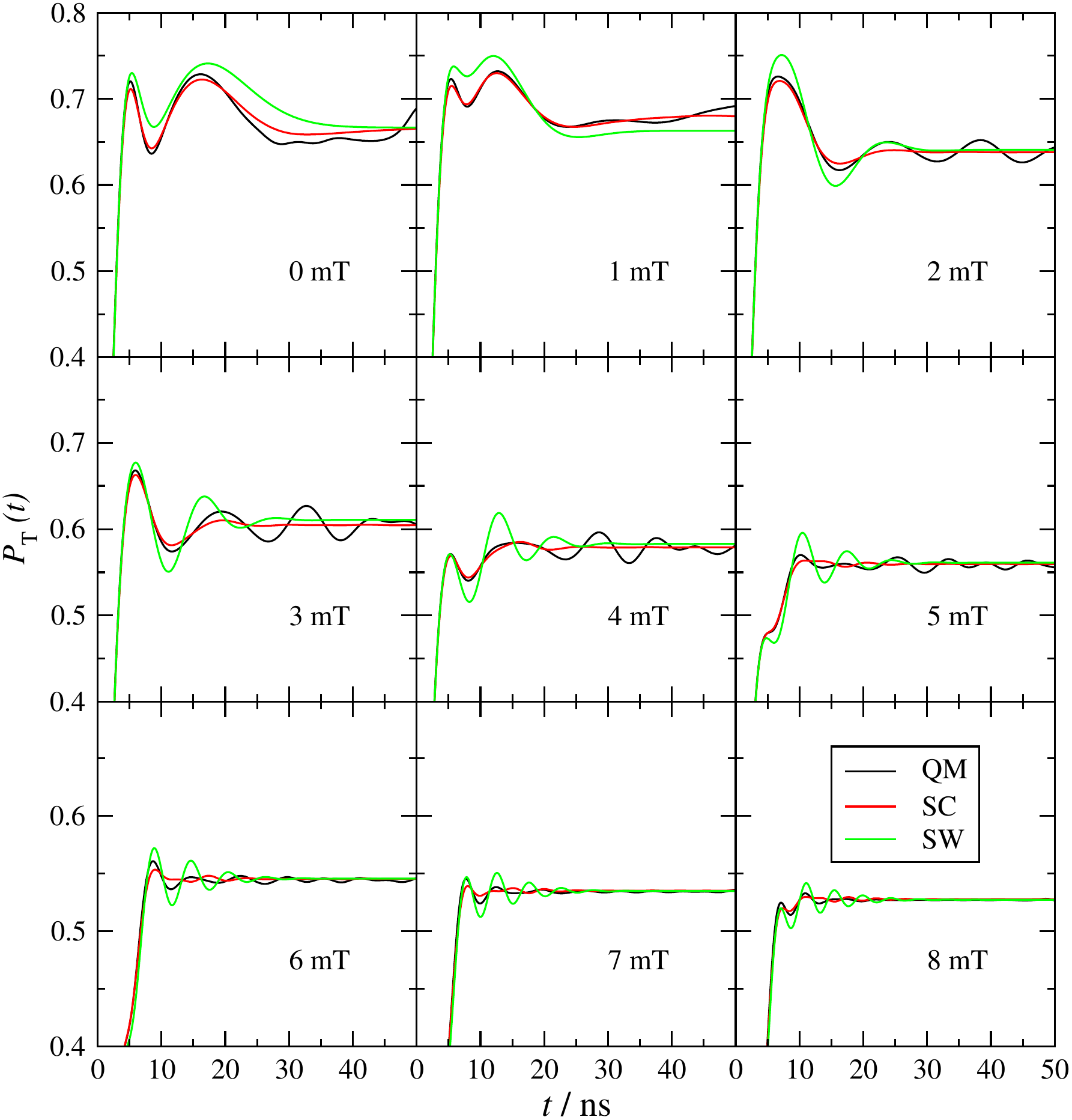}}
\caption{Comparison of quantum mechanical (QM), semiclassical (SC) and Schulten-Wolynes (SW) triplet probabilities for a model of the $^2$Py$^-$+$^2$DMA$^+$ radical pair in various magnetic fields.}
\end{figure}

\begin{figure}[Ht]
\centering
\resizebox{0.8\columnwidth}{!} {\includegraphics{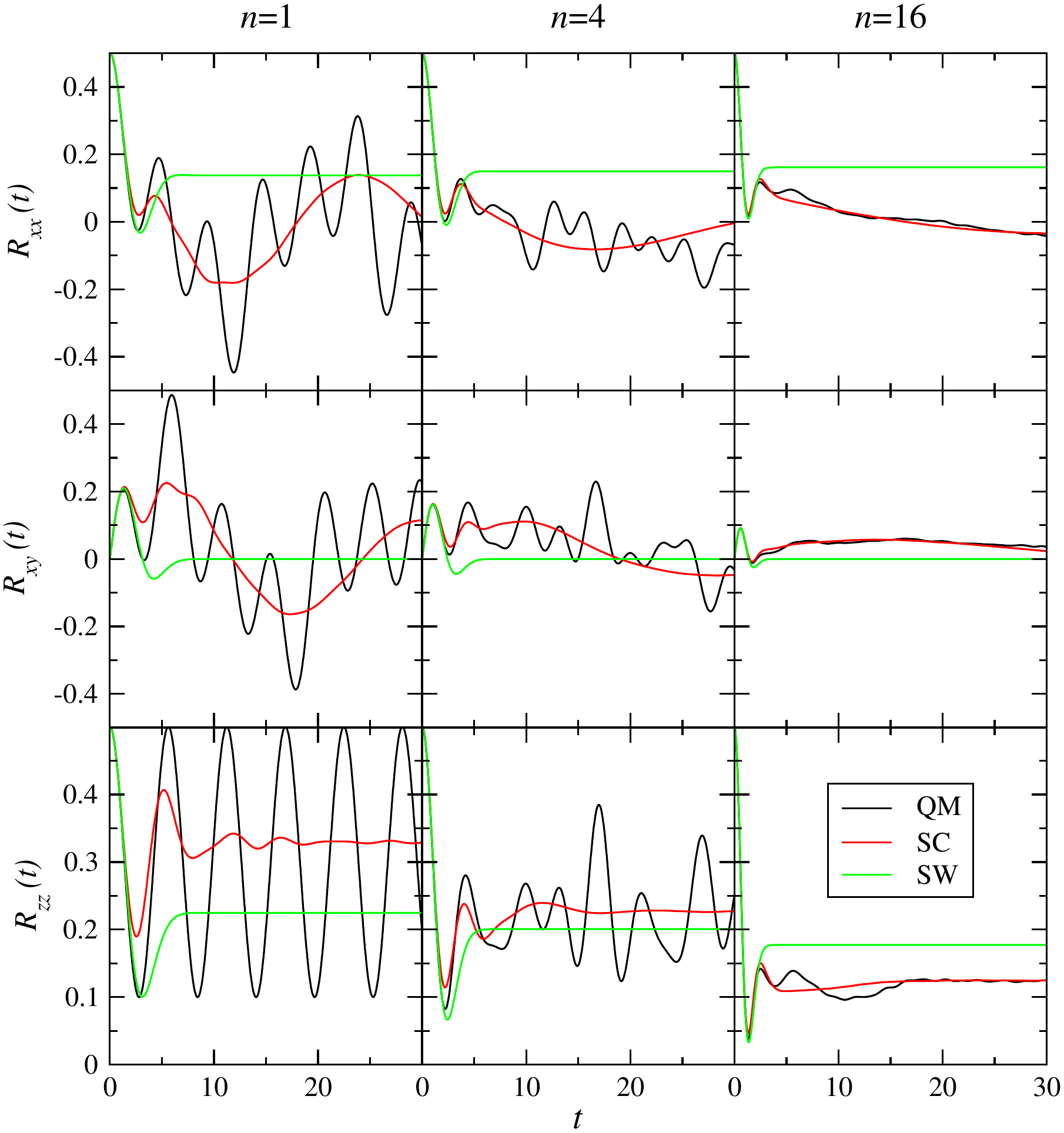}}
\caption{Quantum mechanical (QM), semiclassical (SC), and Schulten-Wolynes (SW) electron spin correlation tensors for radicals with $n=1$, 4 and 16 nuclear spins in a magnetic field of 0.5 mT. The nuclear spins all have $I=1/2$ and their hyperfine couplings to the electron spin are in the range $-1<a_i<1$ mT. The time $t$ is in units of mT$^{-1}$.}
 \end{figure}

\begin{figure}[Ht]
\centering
\resizebox{0.8\columnwidth}{!} {\includegraphics{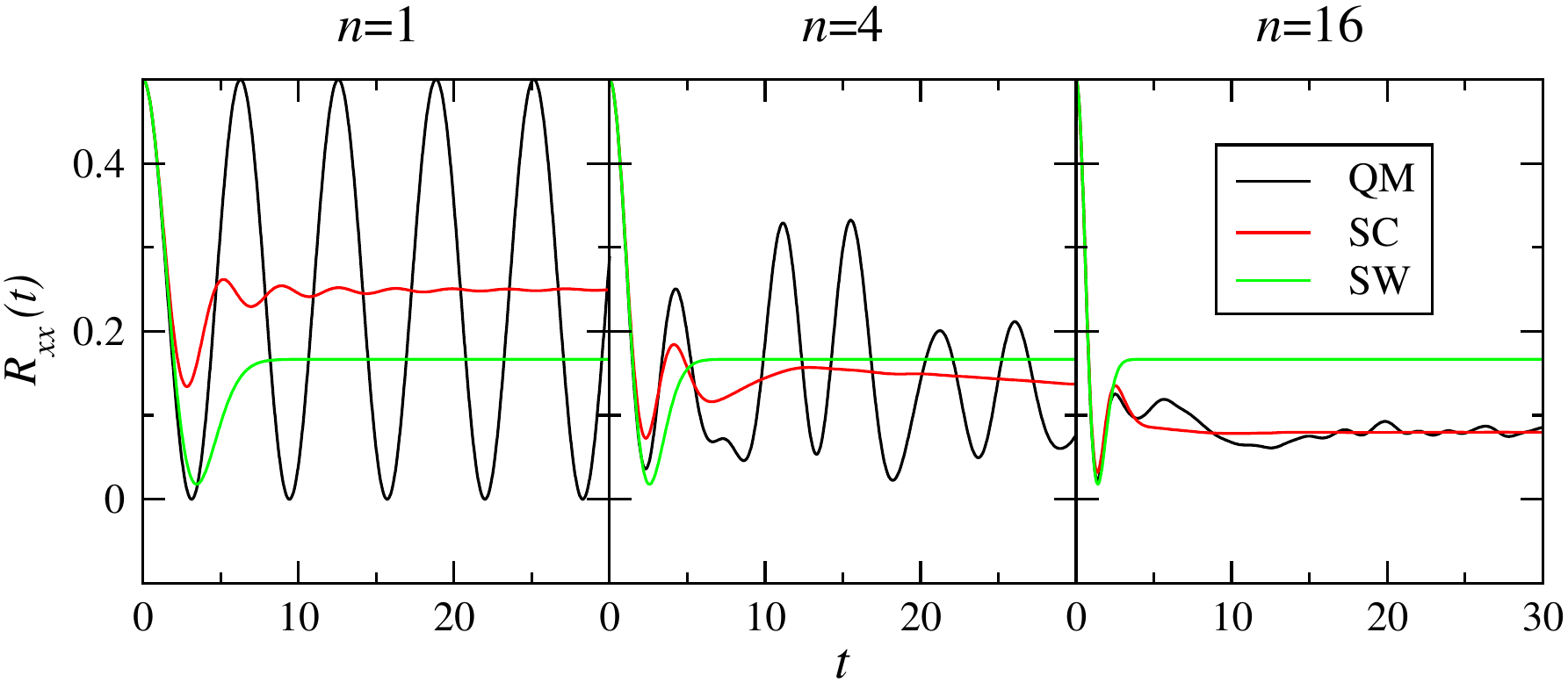}}
\caption{As in Fig.~2, but in the absence of a magnetic field. In this case the electron spin correlation tensor is isotropic, so only $R_{xx}(t)$ is shown.}
\end{figure}

\begin{figure}[Ht]
\centering
\resizebox{0.5\columnwidth}{!} {\includegraphics{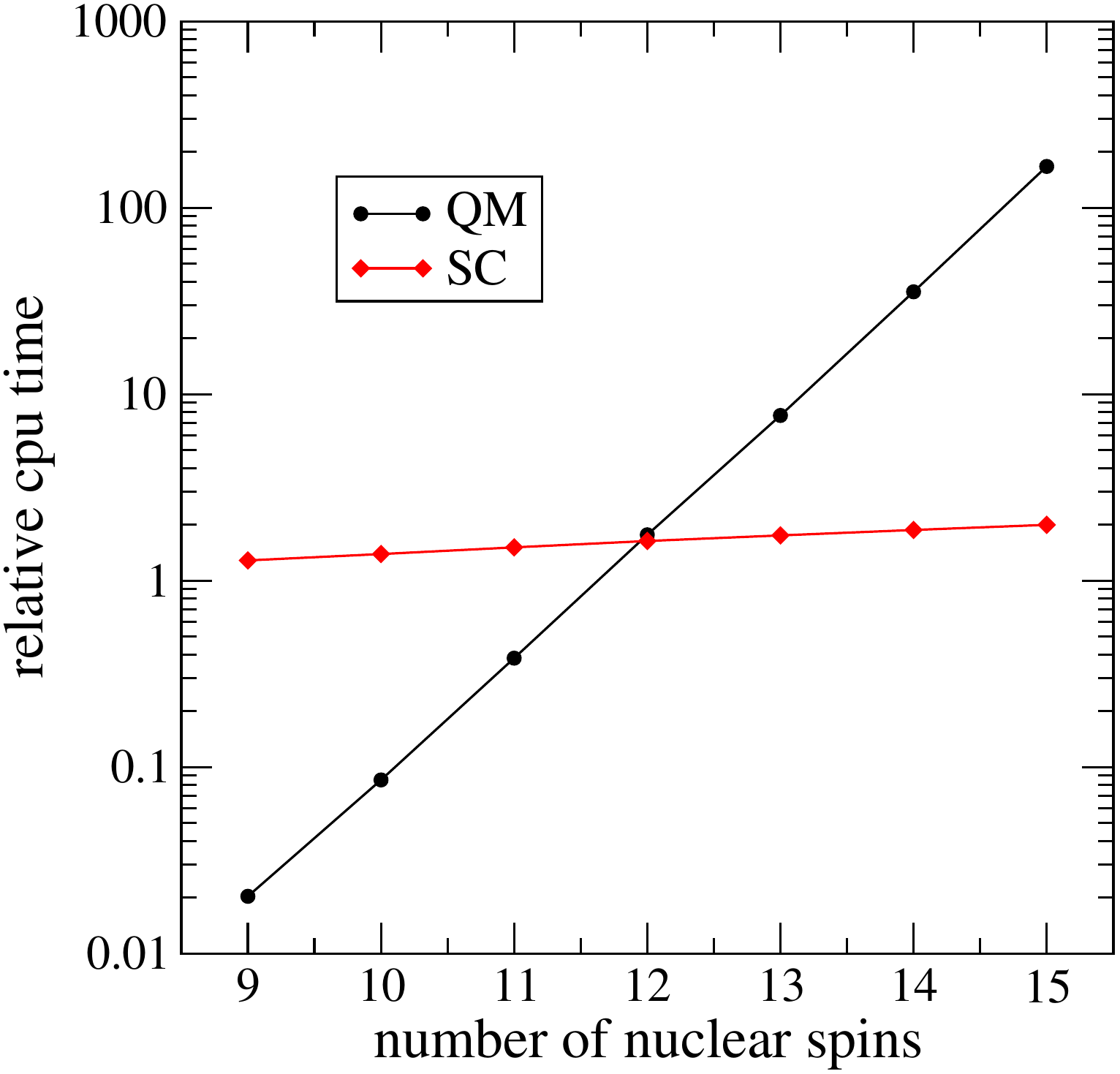}}
\caption{Relative computer times of quantum mechanical (QM) and semiclassical (SC) spin correlation tensor calculations as a function of the number of $I=1/2$ nuclear spins in the radical, obtained using the algorithms detailed in the Appendix. }
\end{figure}
 
\begin{figure}[Ht]
\centering
\resizebox{0.8\columnwidth}{!} {\includegraphics{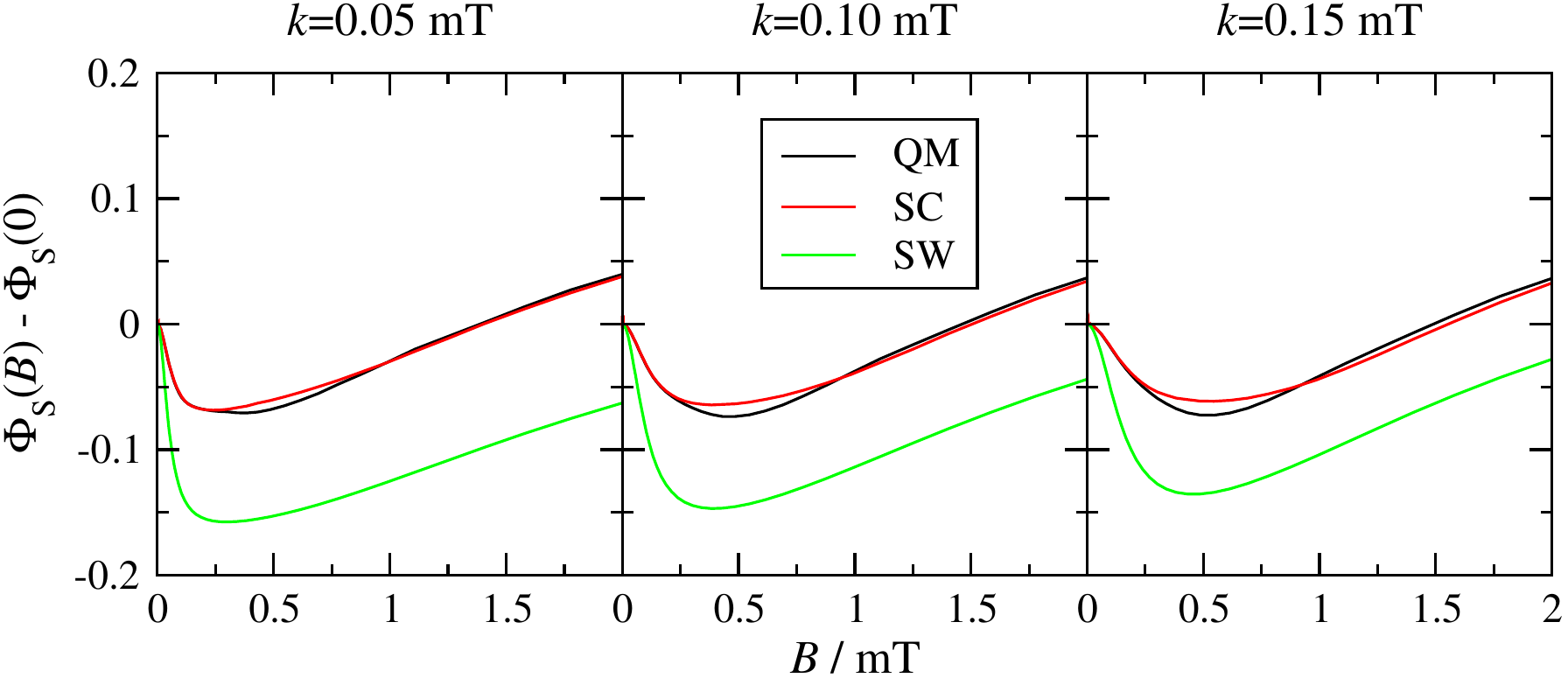}}
\caption{Comparison of low-field QM, SC and SW singlet yields for a radical pair with 12 $I=1/2$ nuclear spins in one radical and none in the other, for three different recombination rates.}
\end{figure}


\begin{thebibliography}{99}
\bibitem{MonteCarlo}
In all of the SC calculations reported in this paper, we averaged over 10$^6$ initial spin orientations, which we checked were enough to give well-converged results by averaging over ten times fewer.
\end{thebibliography}
\end{document}